\documentclass[useAMS,usenatbib]{mn2e}

\usepackage{amsmath, amssymb}
\usepackage{epsfig}         
\usepackage{graphicx, float}
\usepackage{multirow}

\usepackage{natbib}
\usepackage{aas_macros}

\usepackage{url}
\usepackage{hyperref}

\def\Mpch{~h^{-1} {\rm Mpc}}
\def\MpchVolume{~(h^{-1} {\rm Mpc})^3}
\def\kpc{~\rm kpc}
\def\kpch{~h^{-1} {\rm kpc}}
\def\Msun{\rm{M}_{\odot}}

\def\kms{~\rm{km/s}}

\newcommand{\rockstar}{\textsc{rockstar}}

\newcommand{\MII}{\textsc{MS-II}}

\newcommand {\lcdm}{$\Lambda$CDM}

\newcommand{\cum}{({>}\nu)}  
\newcommand{\Ncum}{\overline{N}({>}\nu)}  

\newcommand{\gsim}{\raisebox{-0.3ex}{\mbox{$\stackrel{>}{_\sim} \,$}}}
\newcommand{\lsim}{\raisebox{-0.3ex}{\mbox{$\stackrel{<}{_\sim} \,$}}}
\newcommand{\eq}[1]{Eq. \eqref{#1}}

\newcommand{\refsec}[1]{\S \ref{#1}}
\newcommand{\refappendix}[1]{Appendix \ref{#1}}

\newcommand{\reftab}[1]{Table \ref{#1}}
\newcommand{\reffig}[1]{Fig. \ref{#1}}

\newcommand{\reffigS}[2]{Figs \ref{#1} and \ref{#2}}

\newcommand{\figDir}{fig_pdf/}

\newcommand{\firstPaper}{C14}
\newcommand{\wang}{Wang12}

\usepackage[normalem]{ulem}

\usepackage{color}
\definecolor{colorChanges}{rgb}{.0,.3,1.}

\newcommand{\MCn}[1]{#1} 
\newcommand{\MCd}[1]{} 
\newcommand{\MCc}[1]{} 
\newcommand{\MCq}[1]{} 

\title[Milky Way mass constraints from the Galactic satellite gap]
{Milky Way mass constraints from the Galactic satellite gap}

\author[Cautun~et~al.]
{\parbox{\textwidth}{
        Marius Cautun$^{1,2}$\thanks{E-mail : m.c.cautun@durham.ac.uk},
        Carlos S. Frenk$^{1}$,
        Rien van de Weygaert$^{2}$,
        Wojciech A. Hellwing$^{1,3}$
        and Bernard~J.~T.~Jones$^{2}$\vspace{.4cm}} \\
$^1$   Department of Physics, Institute for Computational Cosmology, University of Durham, South Road Durham DH1 3LE \\
$^2$   Kapteyn Astronomical Institute, University of Groningen, P.O. Box 800, 9747 AV Groningen, The Netherlands \\
$^3$   Interdisciplinary Centre for Mathematical and Computational Modellings, University of Warsaw, ul. Pawi\'nskiego 5a, Warsaw, Poland\\
}

\begin{document}


\maketitle

\begin{abstract}
  We use the distribution of maximum circular velocities, $V_\rmn{max}$,
  of satellites in the Milky Way (MW) to constrain the virial mass, $M_{200}$, of 
  the Galactic halo under an assumed prior of a \lcdm{} universe. 
  This is done by analysing the subhalo populations of a large sample 
  of halos found in the Millennium II cosmological simulation. 
  The observation that the MW has at most three subhalos with $V_\rmn{max}\ge30\kms$
  requires a halo mass $M_{200}\le1.4\times10^{12}\Msun$, while 
  the existence of the Magellanic Clouds (assumed to have $V_\rmn{max}\ge60\kms$)
  requires $M_{200}\ge1.0\times10^{12}\Msun$. The first of these
  conditions is necessary to avoid the ``too-big-to-fail'' problem
  highlighted by Boylan-Kolchin et al., while the second stems from
  the observation that massive satellites like the Magellanic Clouds
  are rare. When combining both requirements, we find that the MW halo mass
  must lie in the range $0.25 \le M_{200}/(10^{12}\Msun) \le 1.4$ 
  at $90\%$ confidence. The gap in the abundance of Galactic 
  satellites between $30\kms\le V_\rmn{max} \le 60\kms$ places
  our galaxy in the tail of the expected satellite distribution.
\end{abstract}



\section{Introduction}
\label{sec:MW_satellites:introduction}
Due to their proximity, the Milky Way (MW) and its satellite galaxies
provide an unparalleled dataset for testing astrophysical and
cosmological ideas. For example, resolving the stellar content of the
dwarf spheroidals enables tests of galaxy formation and evolution
theory \citep{Grebel2005}; analyzing their internal kinematics
constrains the nature of their dark matter content
\citep[e.g.][]{Strigari2010}; detecting satellites three orders of
magnitude fainter than in external galaxies \citep[e.g.][]{Willman05}
provides information on the physics of extreme, very low luminosity
galaxies. Given that the MW satellites play such a prominent role, it
is important to investigate how representative the MW substructures
are of systems of this kind.

Several alleged points of tension between observations and predictions
of the standard cosmological model, \lcdm{}, concern properties of the
MW and its satellites. One is an apparent discrepancy between the
predicted distribution of the maximum circular velocity,
$V_\rmn{max}$, of the most massive subhalos and the inferred values
for the MW satellites. This is often referred to as the ``satellite
problem'', and was originally identified by \citet{Klypin1999} and
\citet{Moore1999}. Another variant of this discrepancy was recently highlighted
by \cite{Parry2012} and by
\cite{Boylan-Kolchin2011a,Boylan-Kolchin2012a} who dubbed it the
``too-big-to-fail'' problem.

Various arguments based on the kinematics of the nine bright
``classical'' dwarf spheroidal satellites of the MW suggest that they reside in
subhalos with maximum circular velocities of $V_\rmn{max}\lsim30\kms$
\citep{Penarrubia2008,Strigari2008,Lokas2009,Walker2009,Wolf2010,Strigari2010},
or even $V_\rmn{max}\lsim25\kms$ \citep{Boylan-Kolchin2012a}.  If this is
indeed the case, only the two Magellanic Clouds (MCs) and the
Sagittarius dwarf would reside in dark matter substructures with
larger maximum velocity than this.  Using the Aquarius simulations
\citep{aquarius2008}, \cite{Boylan-Kolchin2011a,Boylan-Kolchin2012a}
argued that having at most three massive satellites with
$V_\rmn{max}\ge30\kms$ in the MW is in conflict with current
understanding of galaxy formation and evolution within \lcdm{}:
simulations produce, on average, eight, not three, subhalos with
$V_\rmn{max}$ larger than $30\kms$. At face value, this would
require the most massive substructures to be devoid of stars when less
massive objects are not. This is not expected in models of how
galaxies populate low mass halos \citep[e.g.][]{Benson2002} and could
signal a fundamental shortcoming of the \lcdm{} model itself. A
similar conclusion was independently reached by \cite{Parry2012} from
hydrodynamic simulations of galaxy formation in some of the Aquarius
halos.

A possible solution to the ``too-big-to-fail'' (TBTF) problem was put forward
by \citet[hereafter \wang{}]{Wang2012}. Using the approximate
invariance of the scaled subhalo maximum velocity function with host
halo mass \citep[see
e.g.][]{Moore1999,Kravtsov2004,Zheng2005,aquarius2008,Weinberg2008},
\wang{} derived statistics for galactic subhalos and estimated the
probability that a Milky Way halo contains three or fewer satellites
with $V_\rmn{max}\ge30\kms$, as a function of the host halo mass. \MCn{These results were further refined by \citet[][hereafter \firstPaper{}]{Cautun2013a}, who developed a better method for estimating the abundance of galactic subhalos in cosmological simulations.} Both studies
found that rather than ruling out \lcdm{}, the small number of massive
satellites in our galaxy imposes an upper limit to the mass of the MW
halo if \lcdm{} is the correct model. They found that the MW satellite
data are consistent with \lcdm{} predictions at the $10\%$ confidence level if
the MW halo has a virial mass $<1.3\times10^{12}\Msun$, which is near the
lower end of commonly accepted values. \MCn{A similar solution to the TBTF problem was proposed by \citet{Purcell2012}, who compared the structure of MW satellites with that of subhalos predicted by a semi-analytical model. They recognized that the solution to the problem requires the mass of the MW halo to be below a certain value that, however, is significantly larger than the value we find in this paper.}

A low MW halo mass, however, has a large impact on the probability of
finding the two MCs, which are rather massive. Recent estimates
with HST data find maximum circular velocities of $(92\pm19)\kms$ and
$(60\pm5)\kms$ for the Large and Small Magellanic Clouds respectively
\citep{Kallivayalil2013,vanderMarel2014}, which broadly agree with measurements based
on HI and stellar kinematics
\citep[e.g.][]{vanderMarel2002,Stanimirovic2004,Harris2006,Olsen2007}.
Simulation studies agree that, in \lcdm{}, substructures with the
mass of the MCs are common in massive galactic halos, of mass $\sim
2-3\times10^{12}\Msun$, but are quite rare in halos of lower mass,
${\lsim}1\times10^{12}\Msun$
\citep{Boylan-Kolchin2011b,Busha2011b,Busha2011,Gonzalez2013}.  Galaxy
redshift survey data indicate that galaxies with luminosity similar to
the MW have ${\sim}4\%$ probability of hosting two satellites like the
MCs \citep{Liu2011,Guo2011,Lares2011}. Taking into account both mass
and orbital data for the two MCs, \citet{Busha2011b} and
\citet{Gonzalez2013} estimate a mass of ${\sim}1.2\times10^{12}\Msun$
for the MW halo, in contradiction with the conclusion of
\citet{Boylan-Kolchin2011b}, which, using similar considerations, found that the MW
halo mass is unlikely to be less than $2\times10^{12}\Msun$. The
former is consistent with the constraint of \wang{} but the latter is
not.

In this paper we investigate the constraints that the massive
satellite population of the MW sets on the mass of its dark matter
halo in the context of the \lcdm{} model. 
In addition, we remark on the peculiar gap in the number of satellites
in the MW, with at most one satellite in the range $30\kms\le V_{max}\le 60\kms$. 
The TBTF problem is predicated on the basis of this gap. Such gaps 
are rare in our simulations and might signal a tension between the
\lcdm{} model and observations. However, it is not clear how an
{\em a posteriori} argument of this nature can be put on a proper
statistical basis.
This study was possible by making use of a large and representative
sample of simulated halos for which we determine the subhalo number
statistics down to $V_\rmn{max}\sim15\kms$ using the extrapolation
method presented in \firstPaper{}. 

The remainder of this paper is organized as follows. In
\refsec{sec:MW_satellites:data_analysis} we give a description of the
simulations and of the method we employ to extend the dynamic range
over which we derive subhalo count statistics. In
\refsec{sec:MW_satellites:MW_typical} we calculate the probability of
finding MW-like subhalos as a function of halo mass. In
\refsec{sec:MW_satellites:discussion} we examine the sensitivity of
our results to model parameters. We conclude in
\refsec{sec:MW_satellites:conclusion} with a brief summary of our main
results.


\section{The simulations}
\label{sec:MW_satellites:data_analysis}

We make use of the high-resolution Millennium-II cosmological N-body
simulation \citep[\MII{};][]{millSim2}. \MII{} follows the evolution
of cold dark matter, using $2160^3$ particles to resolve
structure formation in a periodic cube $100\Mpch$ on a side. Each 
particle has a mass, $m_p=9.44\times10^6\Msun$, so MW-sized halos
($\sim10^{12}\Msun$) are resolved with $\sim10^5$ particles.  This
represents a good compromise between having a representative sample of
MW-like halos and resolving the most massive $10$ substructures per
host halo. The spatial resolution is given by the Plummer-equivalent
force softening, $\epsilon=1\kpch$, which was kept constant in
comoving coordinates for the entire simulation. \MII{} uses the WMAP-1
cosmogony \citep{Spergel2003} with the following cosmological
parameters: $\Omega_m=0.23$, $\Omega_\Lambda=0.75$, $h=0.73$, $n_s=1$
and $\sigma_8=0.9$.

\subsection{Halo finder}
\label{subsec:MW_satellites:rockstar}
Halos and subhalos in the simulation were identified with the
\rockstar{} (Robust Overdensity Calculation using K-Space
Topologically Adaptive Refinement) phase-space halo finder
\citep{Berhoozi2011}. \rockstar{} starts by selecting potential halos
as Friends-of-Friends (FOF) groups in position space using a large
linking length ($b = 0.28$). This first step is restricted to position
space to optimize the use of computational resources, while each
subsequent step is carried out using the full 6D phase-space
information. Each FOF group from the first step is used to create a
hierarchy of FOF phase-space subgroups by progressively reducing the
linking length. The phase-space subgroups are selected with an
adaptive phase-space linking length such that each successive subgroup
has $70\%$ of the parent's particles. \rockstar{} uses the resulting
subgroups as potential halo and subhalo centres and assigns particles
to them based on their phase-space proximity. Once all particles are
assigned to halos and subhalos, an unbinding procedure is applied to
keep only the gravitationally bound particles. The final halo centres
are computed from a small region around the phase-space density
maximum associated with each object.

The outer boundary of the halos is cut at the point where the
enclosed overdensity decreases below $\Delta=200$ times the critical
density, $\rho_c$. Therefore, the halo mass, $M_{200}$, and radius,
$R_{200}$, correspond to a spherical overdensity of $200\rho_c$. Using
this definition of the main halo boundaries we define the satellite
population as all the subhalos within a distance, $R_{200}$, from the
host centre.

\subsection{Subhalo number statistics}
\label{subsec:MW_satellites:subhalo_statistics}

A challenge when studying galactic substructures in
simulations is to achieve the large dynamic range required for all
subhalos above a certain threshold ($V_\rmn{max}\ge30\kms$ in our
case) to be resolved for a statistically useful sample.  One strategy
is to run ensembles of very high resolution simulations of galactic
halos.
\citep[e.g.][]{Diemand2008,Madau2008,aquarius2008,Stadel2009}.
However, the limited sample size, six in the Aquarius programme, the
largest to date, limits the extent to which they can be used to study
how common the MW satellite systems are. The alternative strategy is
to run simulations of cosmological volumes that produce
representative samples of galactic halos, but are limited in
resolution, so that not all the subhalos above the desired
$V_\rmn{max}$ threshold are resolved \citep{millSim2,Klypin2011}. For
example, while \MII{} captures all substructures with
$V_\rmn{max}\ge45\kms$, it only generates an incomplete population of
less massive subhalos (see \firstPaper{}). To be able to use \MII{}
for our analysis we need to recover the full population of
substructures down to at least $V_\rmn{max}=30\kms$. We now summarize
a procedure introduced in \firstPaper{} for achieving this. 

We are interested in the subhalo abundance as a function of the ratio, 
\begin{equation}
    \nu=\frac{V_\rmn{max}}{V_{200}},
\end{equation} 
between the subhalo maximum velocity, $V_\rmn{max}$, and the virial
velocity, $V_{200}$, of the host halo. We use this quantity to
characterise the halo population because the maximum velocity provides
a robust measurement of subhalo size that is independent of the
identification algorithm and definition of subhalo boundary \citep[for
details see][]{Onions2012}. Moreover, since $V_\rmn{max}$ depends only
on the mass distribution in the central parts of the object, it allows
for a closer comparison with observations that typically probe only
the inner regions of a halo where the galaxy resides.  We now quantify
the statistics of the number of subhalos exceeding $\nu$ and consider
both the mean subhalo count, $\Ncum{}$, and the dispersion,
$\sigma\cum{}$.

\begin{figure}
     \centering
     \includegraphics[width=\linewidth,angle=0]{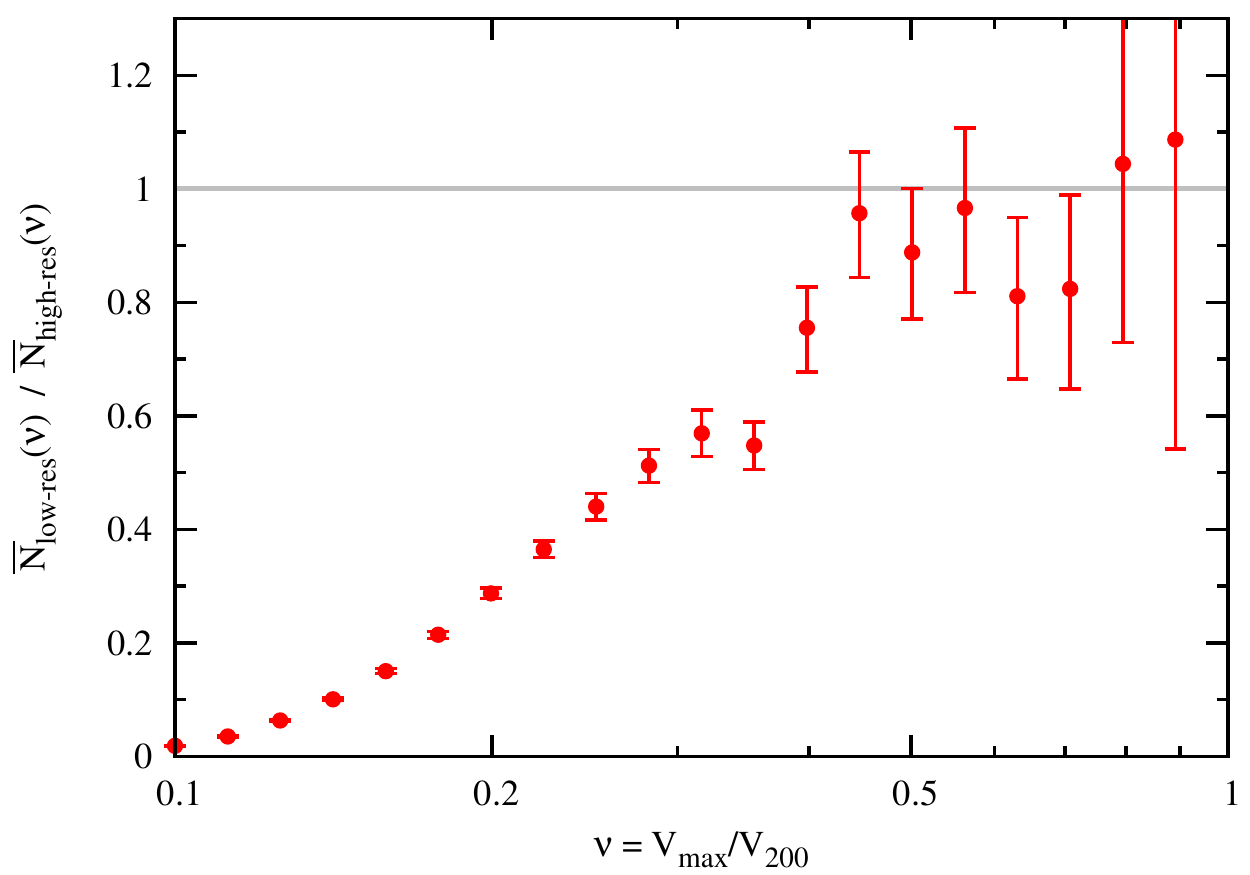}
     \caption{ The impact of numerical resolution on the number of
       subhalos found in simulations. The plot shows the ratio, 
       $\overline{N}_\rmn{low-res}(\nu)/\overline{N}_\rmn{high-res}(\nu)$, 
       between the mean subhalo count in a low and a high
       resolution simulation. A ratio of one corresponds to recovering
       the full substructure population, while lower values reflect 
      missing subhalos in the low resolution simulation. Reproduced from \firstPaper{}. }
     \label{fig:MW_satellites:resolution_effects}
\end{figure}

The effects of limited resolution on the subhalo number counts are
illustrated in \reffig{fig:MW_satellites:resolution_effects}.
It contrasts, as a function of $\nu$, the mean subhalo count 
of $(0.8-1.8)\times10^{13}\Msun$ mass haloes resolved at low 
resolution in the Millennium simulation \citep{millSim} and at $125$ times
 higher mass resolution in the \MII{} (reproduced from \firstPaper{}).
The low resolution calculation recovers the
massive substructures, but only finds a partial population of
subhalos below $\nu\approx0.4$. While the exact value of $\nu$ below
which a given simulation misses subhalos depends on several
parameters, especially the number of particles used to resolve the host halo, the qualitative
behaviour shown in \reffig{fig:MW_satellites:resolution_effects} holds
for a wide range of halo masses.

The subhalo population statistics, $\Ncum{}$ and $\sigma\cum{}$, can
be recovered to up to three times lower values of $\nu$ than is
possible in the simulation itself by using the extrapolation method
described in \firstPaper{}. The first step consists of quantifying how
many substructures are missing at each value of $\nu$ in a given
sample of equal mass halos. Once this is known, the method adds the
missing subhalos using a probabilistic approach. Each new subhalo is
randomly assigned to one of the halos in the sample. This procedure
recovers the subhalo statistics, but not the substructure of
individual halos or their spatial distribution.

By applying our extrapolation method to the \MII{} data, in
\firstPaper{} we studied the subhalo number statistics down to
substructures with $V_\rmn{max}\sim15\kms$. Here we summarise some of
the results of \firstPaper{} that are of importance to the present
study. In \firstPaper{} we have found that the probability distribution
function (PDF) of the number of subhalos exceeding $\nu$ is well 
modelled by a negative binomial distribution \citep[see also][]{Boylan-Kolchin2010}, 
\begin{equation}
	P(N|r,s) = \frac{\Gamma(N+r)}{\Gamma(r)\Gamma(N+1)} s^r (1-s)^N \;,
	\label{eq:MW_satellites:NBD}
\end{equation}
where $\Gamma(x)=(x-1)!$ denotes the Gamma function. The parameters, $r$ and 
$s$, are given in terms of the mean, $\Ncum{}$, and the variance, $\sigma^2\cum{}$, 
of the subhalo population by 
\begin{equation}
    r\cum{} = \frac{\overline{N}^2\cum{}}{\sigma^2\cum{}-\Ncum} \mbox{\hspace{.2cm} and \hspace{.2cm}} s\cum{} = \frac{\Ncum}{\sigma^2\cum}
    \label{eq:MW_satellites:NBD_parameters} \;.
\end{equation}
To obtain the substructure number distribution functions, we employ
the mean and the dispersion of the subhalo population computed in
\firstPaper{}. While in \firstPaper{} these quantities were computed
for halos in the mass range $(0.8-3)\times10^{12}\Msun$, the results
are largely independent of the exact halo mass (see \firstPaper{} and 
\reffig{fig:statistics_mass_dependence}).


\section{Limits on the Milky Way halo mass}
\label{sec:MW_halo_mass}

In this section we use the subhalo statistics of galactic halos to 
constrain the mass of the MW halo assuming the \lcdm{} model.
As we discussed in the introduction, various studies suggest that in
the MW only the two MCs and the Sagittarius dwarf reside in halos of
maximum circular velocity, $V_\rmn{max}\ge30\kms$. HI and stellar
kinematics data suggest that the subhalos of the MCs have
$V_\rmn{max}\ge60\kms$ \citep[][]{Kallivayalil2013}. Therefore, the MW
has at most three subhalos with $V_\rmn{max}\ge30\kms$ and at least
two with $V_\rmn{max}\ge60\kms$. We denote such a population of
substructures as a {\it MW-like subhalo system}.

We first obtain the fraction of halos containing three or fewer
subhalos with $V_\rmn{max}\ge30\kms$ in the \lcdm{} model and, following
\wang{}, use this to set an upper limit to the MW halo mass. We then
independently obtain the probability that a halo has at least two
substructures with $V_\rmn{max}\ge60\kms$ and set a lower limit on the
MW halo mass. 

\begin{figure}
     \centering
     \includegraphics[width=\linewidth,angle=0]{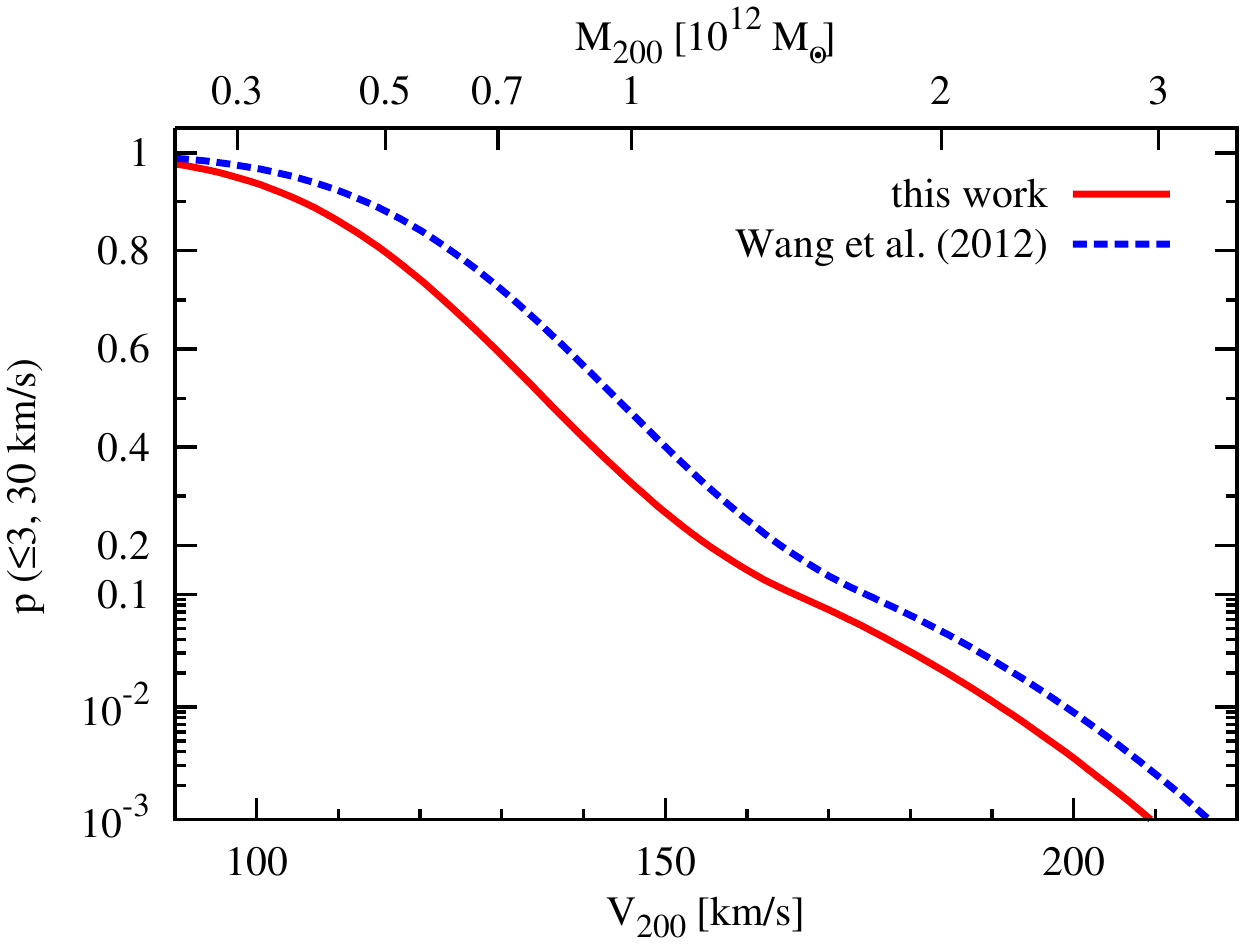}
     \caption{ The probability, $p({\le}3,30\kms)$, that a halo contains
       at most three subhalos with $V_\rmn{max}\ge30 \kms$ as a
       function of the host virial velocity, $V_{200}$, (lower tick marks)
       and virial mass, $M_{200}$, (upper tick marks). The solid
       curve gives our results, while the dashed line shows the
       previous results of \wang{}. Note that the y-axis is linear
       above 0.1 and logarithmic for lower values.}
     \label{fig:MW_constraints_1}
\end{figure}

\subsection{An upper limit to the Milky Way halo mass}
\label{subsec:upper_limit} 

The negative binomial distribution, $P(k|r({>}\nu_0),s({>}\nu_0))$, of
Eq. \ref{eq:MW_satellites:NBD} gives the PDF that a halo has $k$
subhalos with velocity ratio exceeding $\nu_0\equiv V_0/V_{200}$.  It
is then straightforward to estimate the probability that a halo has at
\textbf{most} $X$ substructures with $V_\rmn{max}\ge V_0$. This is
simply the fraction of halos that have at most $X$ subhalos with
$\nu\ge\nu_0$ and can be obtained by summing over the subhalo
abundance PDF at $\nu_0$:
  \begin{equation}
    p({\le}X,V_0) = \sum_{k=0}^{X}  P(k|r({>}\nu_0),s({>}\nu_0)) \mbox{\hspace{.2cm} with} \;\nu_0=\frac{V_0}{V_{200}} \,.
    \label{eq:fraction_le}
\end{equation}
The distribution parameters, $r\cum{}$ and $s\cum{}$, are uniquely
determined by the mean $\Ncum{}$ and scatter $\sigma\cum{}$ of the
subhalo population via \eq{eq:MW_satellites:NBD_parameters}.

The fraction of galactic halos, $p({\le}3,30\kms)$, with at most
three subhalos with $V_\rmn{max}\ge30\kms$ is given in
\reffig{fig:MW_constraints_1} as a function of the host virial
velocity, $V_{200}$ (lower tick marks), and, equivalently, host virial
mass, $M_{200}$ (upper tick marks). For clarity, we plot the halo
fraction on a linear scale for values larger than $0.1$ and on a
logarithmic scale for smaller values. The probability of having at
most three subhalos with $V_\rmn{max}\ge30\kms$, shown as a thick red
curve, is a steep function of host mass, decreasing from $33\%$ at
$10^{12}\Msun$ to $0.1\%$ at $3\times10^{12}\Msun$. For convenience,
we summarize the probabilities for indicative halo masses in
\reftab{tab:MW_probability}. Under the assumption that \lcdm{} is the
correct model, our results then imply a $90\%$ confidence upper limit of
$1.4\times10^{12}\Msun$ for the virial mass of the MW halo, $M_{200}$; a mass of
$2\times10^{12}\Msun$ is ruled out at $97.7\%$ confidence.

The probability of finding at most three halos with
$V_\rmn{max}\ge30\kms$ as a function of $V_{200}$ was previously
derived by \wang{} whose results are shown by the dashed curve in
\reffig{fig:MW_constraints_1}. We find slightly lower upper limits
than them for the mass of the MW halo because they underestimated the
subhalo mass at which resolution effects become important. As a
result, they found $20\%$ fewer substructures than we do (see
\firstPaper{} for more details), causing them to overestimate
$p({\le}3,30\kms)$ at a given halo mass.

\begin{figure}
     \centering
     \includegraphics[width=\linewidth,angle=0]{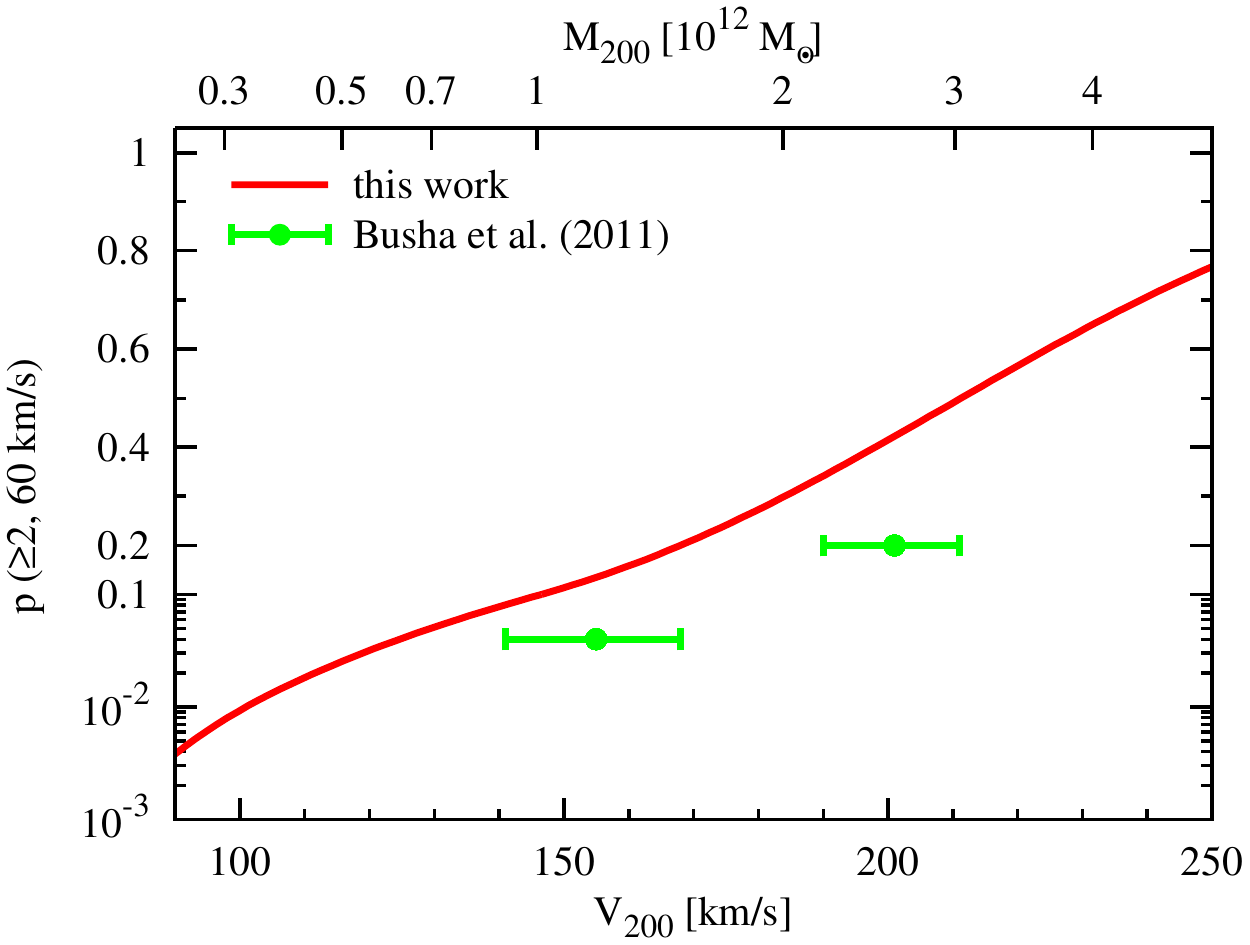}
     \caption{The probability, $p({\ge}2,60\kms)$, that a halo
       contains at least two subhalos with $V_\rmn{max}\ge60 \kms$ as
       a function of the host virial velocity, $V_{200}$, (lower axis),
       and virial mass, $M_{200}$, (upper axis). The solid curve shows
       our predictions, while the filled circles show the results of
       \citet{Busha2011}. Note that the y-axis is linear above 0.1 and
       logarithmic for lower values.}
     \label{fig:MW_constraints_2}
\end{figure}

\begin{table}
    \small
    \centering
    \caption{The fraction of \MII{} halos with massive subhalos similar 
    to those of the MW. The table lists the probability, $p({\le}3,30\kms)$, 
    of finding at most three subhalos with $V_\rmn{max}\ge30\kms$, and 
    the probability, $p({\ge}2,60\kms)$, of finding at least two subhalos 
    with $V_\rmn{max}\ge60\kms$. The last row gives the combined probability 
    of satisfying both conditions simultaneously. } 
    \label{tab:MW_probability}
    \begin{tabular}{lccccc}
        \hline \hline
        Halo mass & $[\times 10^{12}\Msun ]$ & \parbox[][][c]{0.02\textwidth}{\centering $0.5$} & \parbox[][][c]{0.02\textwidth}{\centering $0.7$}   & \parbox[][][c]{0.02\textwidth}{\centering $1$}   & \parbox[][][c]{0.02\textwidth}{\centering $2$}   \\
        \hline
        \parbox[][][c]{0.15\textwidth}{$p({\le}3,30\kms)$}          & $[\%]$ &  80 & 59  & 33 &  2.3 \\[.1cm]
        \parbox[][][c]{0.15\textwidth}{$p({\ge}2,60\kms)$}          & $[\%]$ & 2.2 & 4.7 & 10 &   30 \\[.1cm]
        \parbox[][][c]{0.155\textwidth}{$p({\ge}2,60\kms; {\le}3,30\kms)$}                    & $[\%]$ & 0.8 & 0.7 & 0.4 & 0.04 \\
        \hline
    \end{tabular}
\end{table}

\subsection{A lower limit to the Milky Way halo mass}
\label{subsec:lower_limit} 

The fraction of halos which have at \textbf{least} $X$ subhalos with
$V_\rmn{max}\ge V_0$ can be expressed as
\begin{equation}
    p({\ge}X,V_0) = 1 - p({\le}X{-}1,V_0) \,,
    \label{eq:fraction_ge}
\end{equation}
with $p({\le}X{-}1,V_0)$ given by \eq{eq:fraction_le}.

The probability, $p({\ge}2,60\kms)$, of a halo hosting at least two
subhalos with $V_\rmn{max}\ge60\kms$ is shown as a solid curve in
\reffig{fig:MW_constraints_2}. This represents the fraction of halos
that host MCs-like or more massive substructures as a function of the
$V_{200}$ or $M_{200}$ of the host halo. This probability is small in
low mass halos but increases rapidly towards more massive
hosts. Therefore, assuming \lcdm{}, $p({\ge}2,60\kms)$ sets a lower
limit on the MW halo mass. From \reffig{fig:MW_constraints_2}, we
find a lower limit of $1.0\times 10^{12}\Msun$ for the mass of the
MW halo at 90\% confidence.

The probability of finding two or more substructures with
$V_\rmn{max}\ge60\kms$ in galactic halos was previously estimated by
\cite{Busha2011} whose results are shown as filled circles in
\reffig{fig:MW_constraints_2}. Our values are a factor of a few higher
than theirs. We suspect that the difference arises because
\cite{Busha2011} used the Bolshoi simulation \citep{Klypin2011} which
misses a large number of MCs-like substructures due to numerical
resolution effects. Bolshoi has approximatively the same number of
dark matter particles as \MII{}, but a volume ${\sim}15$ times
larger. Given that \MII{} misses subhalos with $V_\rmn{max}<45\kms$
(see \firstPaper{}), we suspect that the Bolshoi simulation
underestimates the number of substructures with $V_\rmn{max}$ below
$45\kms \times 15^{1/3}\sim100\kms$.

\section{ The mass distribution of the MW }
\label{sec:MW_satellites:MW_typical}

In this section we estimate the mass of the MW, given that our galaxy contains at most three subhalos with $V_\rmn{max}\ge30\kms$, out of which two have at least $V_\rmn{max}\ge60\kms$, to which we refer as a MW-like subhalo system. A crucial ingredient of this analysis is the correlation between the presence of satellites with $V_\rmn{max}\ge60\kms$ and those with $V_\rmn{max}\ge30\kms$, which we estimate from cosmological simulations. This is in contrast to the results of the previous section which treated the two satellite populations as independent, which is clearly not the case.

\begin{figure}
     \centering
     \includegraphics[width=\linewidth,angle=0]{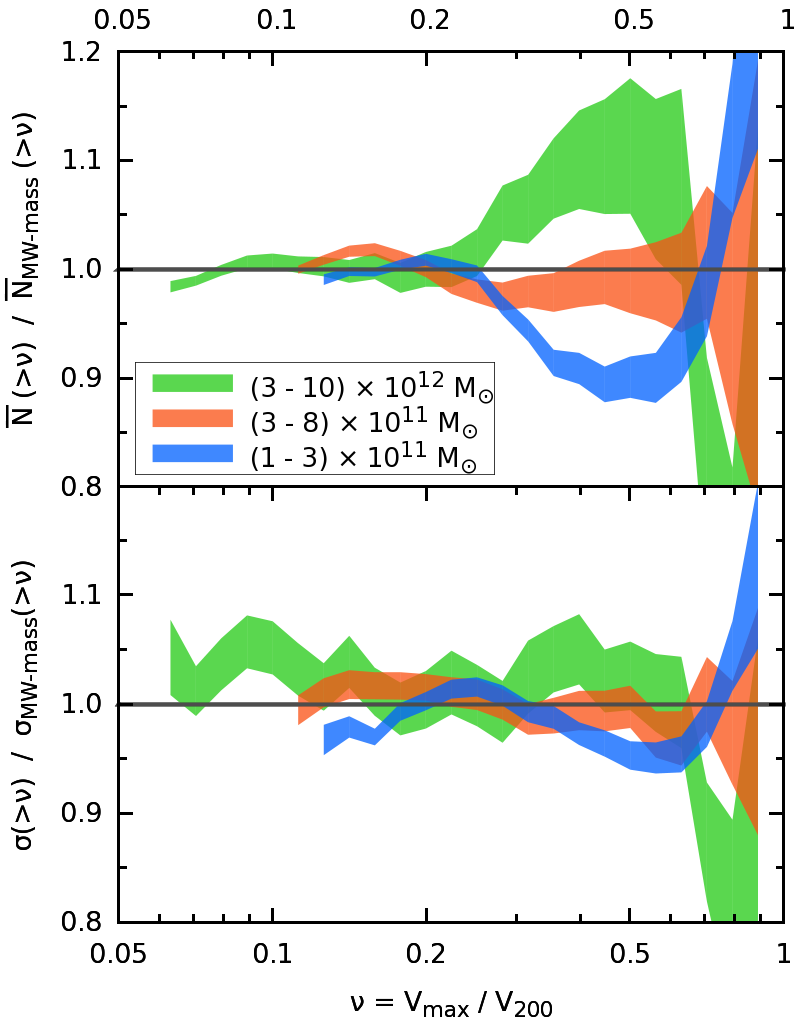}
     \caption{The mean, $\Ncum{}$ (top panel), and the dispersion,
       $\sigma\cum{}$ (lower panel), of the subhalo abundance as a
       function of velocity ratio, $\nu$, for halos in different mass
       bins. For clarity, we plot the ratio with respect to the values
       for halos in the mass range $(0.8-3)\times10^{12}\Msun$. A
       ratio of $1$ corresponds to no variation with host mass. The
       vertical width of the curves shows the bootstrap error associated with $\Ncum{}$ and $\sigma\cum{}$.
      }
     \label{fig:statistics_mass_dependence}
\end{figure}

To obtain the mass distribution of haloes that contain MW-like satellite systems, we compute the probability,
$p({\ge}X_1,V_1;\,{\le}X_2,V_2)$, that a halo contains at least $X_1$
subhalos with $V_\rmn{max}\ge V_1$ and at most $X_2$ substructures
with $V_\rmn{max}\ge V_2$.
As we shall see later, this probability is
quite small for the kind of MW subhalos of interest here and thus a
large sample of halos is required for a robust estimate. Due to its
limited volume, the \MII{} does not provide sufficient statistics for
galactic halos.

Following \wang{}, we can overcome this limitation by appealing to the
approximate invariance of the scaled subhalo velocity function,
$\Ncum{}$, with host halo mass, that is, to the fact that, to good
approximation, the subhalo number PDF is independent of halo mass when
expressed as a function of $\nu$ \citep[][\wang{},
\firstPaper{}]{Moore1999,Kravtsov2004,Zheng2005,aquarius2008,Weinberg2008}.
This is clearly seen in \reffig{fig:statistics_mass_dependence} which
compares the mean and the dispersion of the subhalo number counts in
halos of different mass. We take halos in the mass range
$(0.8-3)\times10^{12}\Msun$ as reference since this interval
encompasses the likely value for the MW as seen in the preceding
section and also as argued by e.g.
\citet{Battaglia2005,Dehnen2006,Xue2008,Gnedin2010,Guo2010}. The
figure shows that, to (10-20)\% accuracy, the number of substructures
is independent of host halo mass over the mass range $10^{11}\Msun -
10^{13}\Msun$.

To proceed further, we rewrite the probability in terms of constraints
on the velocity ratio, $\nu$. Given a halo of virial velocity, 
$V_{200}$, we define
\begin{equation}
    \nu_1 = \frac{V_1}{V_{200}} \mbox{\hspace{0.5cm} and \hspace{0.5cm}} 
    \nu_2 = \frac{V_2}{V_{200}} \;.
    \label{eq:MW_satellites:nu_1_nu_2}
\end{equation}
Computing $p({\ge}X_1,V_1;\;{\le}X_2,V_2)$ now reduces to finding the
probability that a halo contains at least $X_1$ subhalos with $\nu\ge
\nu_1$ and at most $X_2$ subhalos with $\nu\ge \nu_2$. 

\begin{figure}
     \centering
     \includegraphics[width=\linewidth,angle=0]{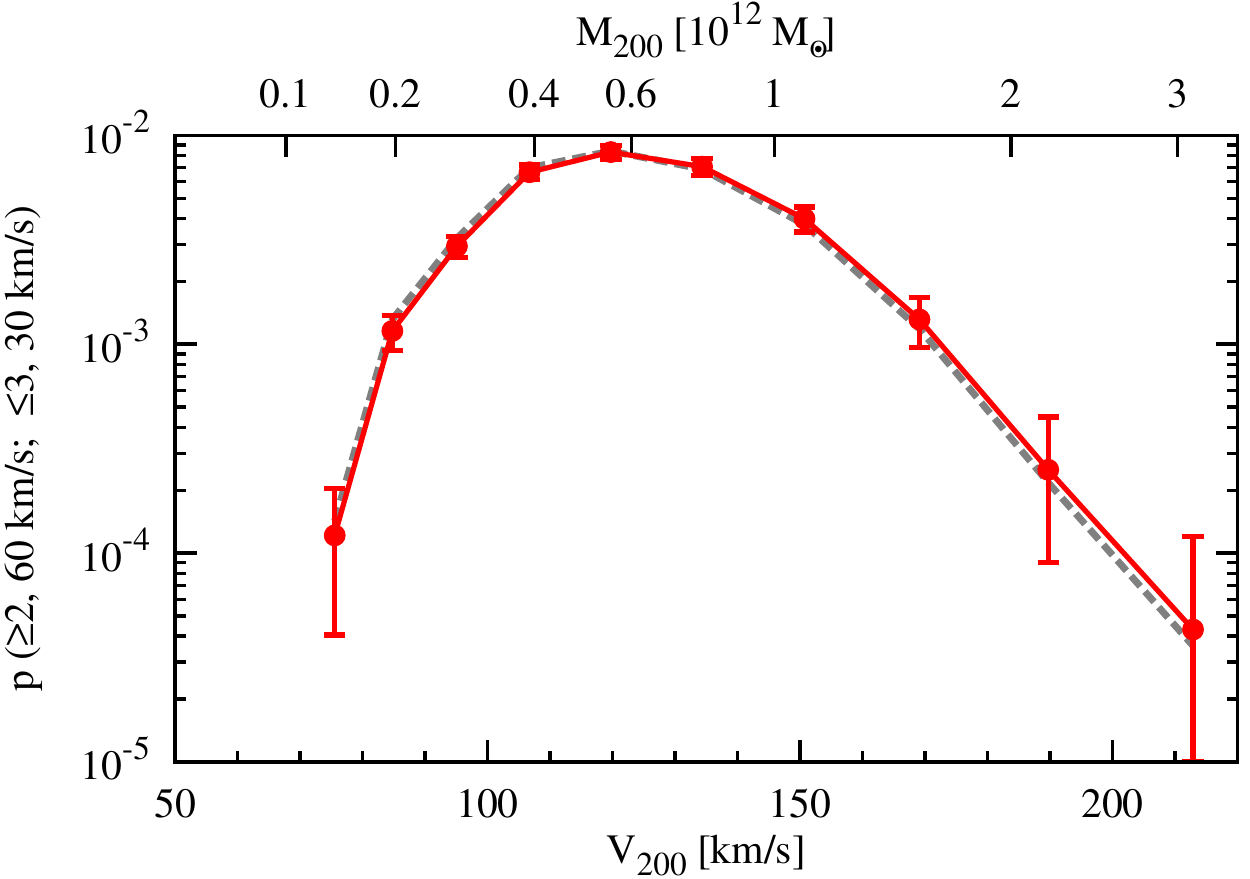}
     \caption{The probability, $p({\ge}2,60\kms;\;{\le}3,30\kms)$,
       that a halo has a MW-like subhalo population as a function of
       halo virial velocity (lower tick marks) and virial mass
       (upper tick marks). The error bars show the 1$\sigma$ spread
       due to the finite number of halos and different realisations
       of the subhalo extrapolation method. 
       The dashed grey line shows the size of the shift towards 
       lower $V_{200}$ values when multiplying the probability by the halo 
       mass function.
       Note the logarithmic y-axis. }
     \label{fig:MW_like_subhalos}
\end{figure}

The probability of finding a MW-like substructure population in the
\MII{} is given in \reffig{fig:MW_like_subhalos} as a function of
both halo virial velocity and halo mass. The probability has a peak
value of ${\sim}1\%$, i.e. at most one out of 100 halos of that
mass has a MW-like subhalo population. Thus, satellite systems such as
the one in our galaxy are rare in a \lcdm{} universe.

The rarity of the MW subhalo population depends strongly on the mass
of the MW halo. The probability is largest for halos in the mass range $\sim
(0.4-1.0)\times10^{12}\Msun$ and drops off sharply outside this
interval, decreasing below one tenth of its peak value outside the
mass range $(0.2-1.5)\times10^{12}\Msun$. 

\MCn{To constrain the MW halo mass we need to multiply the probability of finding a MW-like subhalo system in a halo of a given mass, $p({\ge}2,60\kms;\;{\le}3,30\kms)$, by the total number of halos of that mass. This gives the mass distribution of haloes with MW-like satellite systems}\footnote{\MCn{This is equivalent to taking a flat prior over halo masses, which is the simplest prior to assume}.}.
Due to the sharp drop of the probability outside its peak, multiplying by the halo mass function results only in a slight shift of the distribution to lower halo masses. This is shown by the dashed grey line in \reffig{fig:MW_like_subhalos}. This shift is negligible in comparison to other uncertainties, as we discuss in \refsec{sec:MW_satellites:discussion}, and, to a good approximation, can be neglected.

To obtain the new MW mass constraints, we identify the region under the $p({\ge}2,60\kms;\;{\le}3,30\kms)$ curve that contains $90\%$ of the area. This gives a MW mass range of $(0.25-1.4)\times10^{12}\Msun$, at $90\%$ confidence, with a most likely value of $0.6\times10^{12}\Msun$ given by the peak of the distribution. While the upper limit is the same as we found earlier using the halo fraction, $p({\le}3,30\kms)$, the lower mass limit is significantly lower than the $1.0\times10^{12}\Msun$ value inferred from the $p({\ge}2,60\kms)$ analysis. Thus, treating the MW satellite numbers with $V_\rmn{max}\ge60\kms$ and $V_\rmn{max}\ge30\kms$ independently of each other gives a Galactic mass range that is both narrower and centred at larger values.

\begin{figure}
     \centering
     \includegraphics[width=.86\linewidth,angle=0]{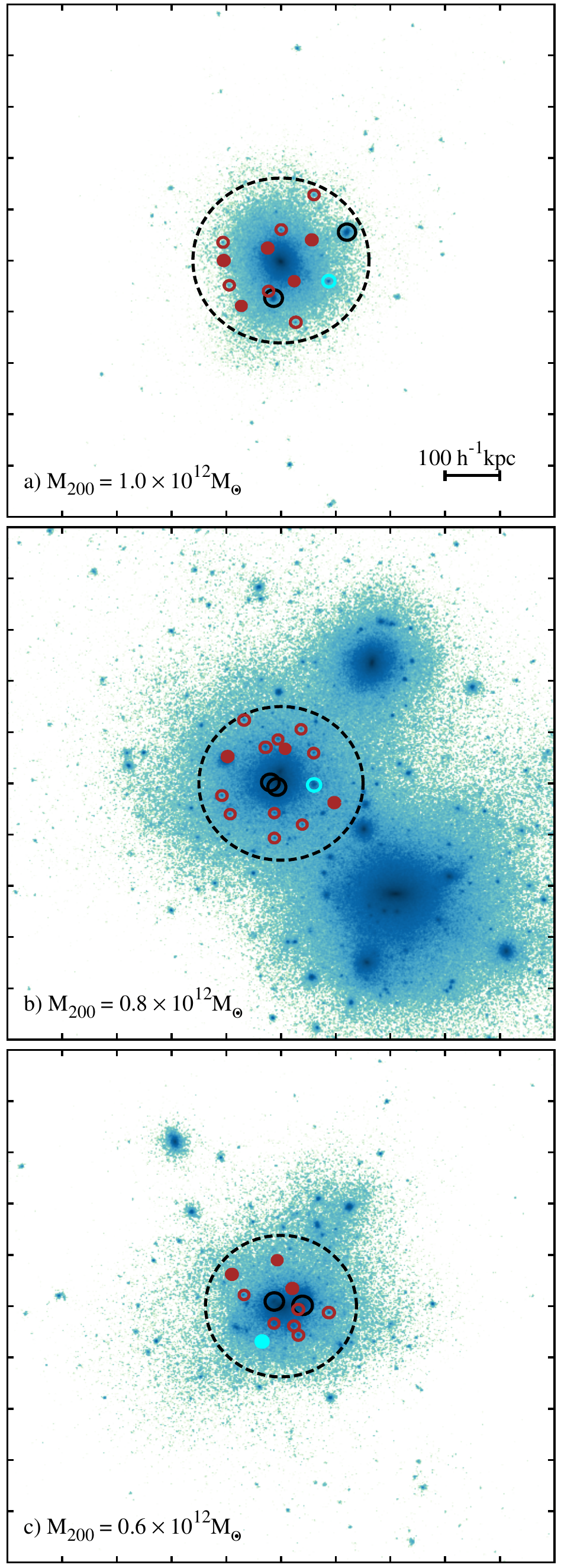}
     \caption{ Examples of \MII{} halos that have a similar subhalo
       population to the MW. Each panel shows a $1\times 1\times
       0.5\MpchVolume$ projection centred on the halo. The black
       dashed circle indicates the virial radius.  The solid circles
       inside the virial radius mark substructures with:
       $V_\rmn{max}\ge60\kms$ (black), $30\kms\le
       V_\rmn{max}\le60\kms$ (cyan) and $20\kms\le
       V_\rmn{max}\le30\kms$ (dark red). The empty circles correspond
       to subhalos found in the simulation, while the filled circles
       correspond to subhalos added by our extrapolation method to
       compensate for numerical resolution effects. }
     \label{fig:halo_example}
\end{figure}

In \reffig{fig:halo_example} we illustrate a few examples of halos
that could potentially contain a MW-like subhalo population\footnote{
These halos correspond to one realization of the subhalo
extrapolation method. Since the method includes a random element, it
cannot recover the substructures of an individual halo and so we can
only identify potential candidates.}. 
We find candidate halos with
a wide range of masses and embedded in a variety of large scale 
environments. For example, the halos in panels a) and c) do not have
similarly massive neighbours in their vicinity, while the halo in
panel b) is part of a group with at least one more massive member.
Substructures with $V_\rmn{max}\ge20\kms$ found within the virial
radius of each object are marked with solid circles. Even though each
of the four halos has at most three massive satellites, they contain
tens of subhalos with $20\kms\le V_\rmn{max}\le30\kms$ that can host the
MW dwarf spheroidal satellites.

\subsection{A model for the probability of having a MW-like subhalo population}
\label{subsec:model_MW_like_subhalos}

In this section we introduce a theoretical model that makes use of
subhalo population statistics to predict the probability that a halo
contains a population of substructures similar to that of our
galaxy. This model is useful for exploring how the conclusions of the
previous section depend on the assumed values of its parameters.

For example, given that at most $1\%$ of halos at any mass have
MW-like subhalos, investigating $p({\ge}2,60\kms;\;{\le}3,30\kms)$
for a different cosmological model requires the analysis of
${\sim}10^4$ MW-mass halos and their substructures, which is a
considerable computational effort. In contrast, obtaining robust
subhalo population statistics can be done using a smaller number of
halos, and therefore the same outcome can be obtained much faster and
cheaper.

We are interested in an analytical model that describes the
probability for a halo to contain at least 2 substructures with
$\nu\ge\nu_1$ and at most 3 substructures with $\nu\ge\nu_2$. The only
hosts that contribute to this probability are those that have:
\begin{itemize}
    \item 2 subhalos with $\nu\ge\nu_1$ and 0 or 1 with $\nu\in[\nu_2,\nu_1]$ or
    \item 3 subhalos with $\nu\ge\nu_1$ and 0 with $\nu\in[\nu_2,\nu_1]$.
\end{itemize}
Assuming that the number of subhalos in the interval $[\nu_2,\nu_1]$
is independent of the the number of subhalos above $\nu_1$, the
contribution of each of the above two terms is given by:
\begin{equation}
    P(k|r({>}\nu_1),s({>}\nu_1)) \times P_\rmn{Poisson}(\le l) \;.
\end{equation}
The first part of the equation is the negative binomial distribution
that gives the fraction of halos that contain $k$ subhalos with
$\nu\ge\nu_1$ (see Eq. \ref{eq:MW_satellites:NBD}). The second part is
the probability that a host contains at most $l$ subhalos in the
interval $[\nu_2,\nu_1]$. This we model using a Poisson distribution,
$P_\rmn{Poisson}(\le l)$. In the range $[\nu_2,\nu_1]$ each halo
contains on average
\begin{equation}
    \Delta N = \overline{N}({>}\nu_2) - \overline{N}({>}\nu_1)
\end{equation}
subhalos. Assuming that this number follows a Poisson distribution
with mean $\Delta N$, the probability that a halo has $l$ subhalos in
the interval $[\nu_2,\nu_1]$ is given by,
\begin{equation}
    \frac{\Delta N^l}{l!}e^{-\Delta N} \;.
    \label{eq:MW_satellites:Poisson_1}
\end{equation}
Putting everything together, we obtain the probability,
$p({\ge}2,60\kms;\;{\le}3,30\kms)$, of finding a halo with a subhalo
population similar to that in the MW, which is given by
\begin{equation}
     \sum_{k=2}^{3} P(k|r({>}\nu_1),s({>}\nu_1)) \;\; \sum_{l=0}^{3-k}\frac{\Delta N^l}{l!} e^{-\Delta N} \;.
\end{equation}
We refer to \refappendix{appendix_model_MW_like_subhalos} for a
derivation of the model and its predictions for the more general case
of $p({\ge}X_1,V_1;\;{\le}X_2,V_2)$.

\begin{figure}
     \centering
     \includegraphics[width=\linewidth,angle=0]{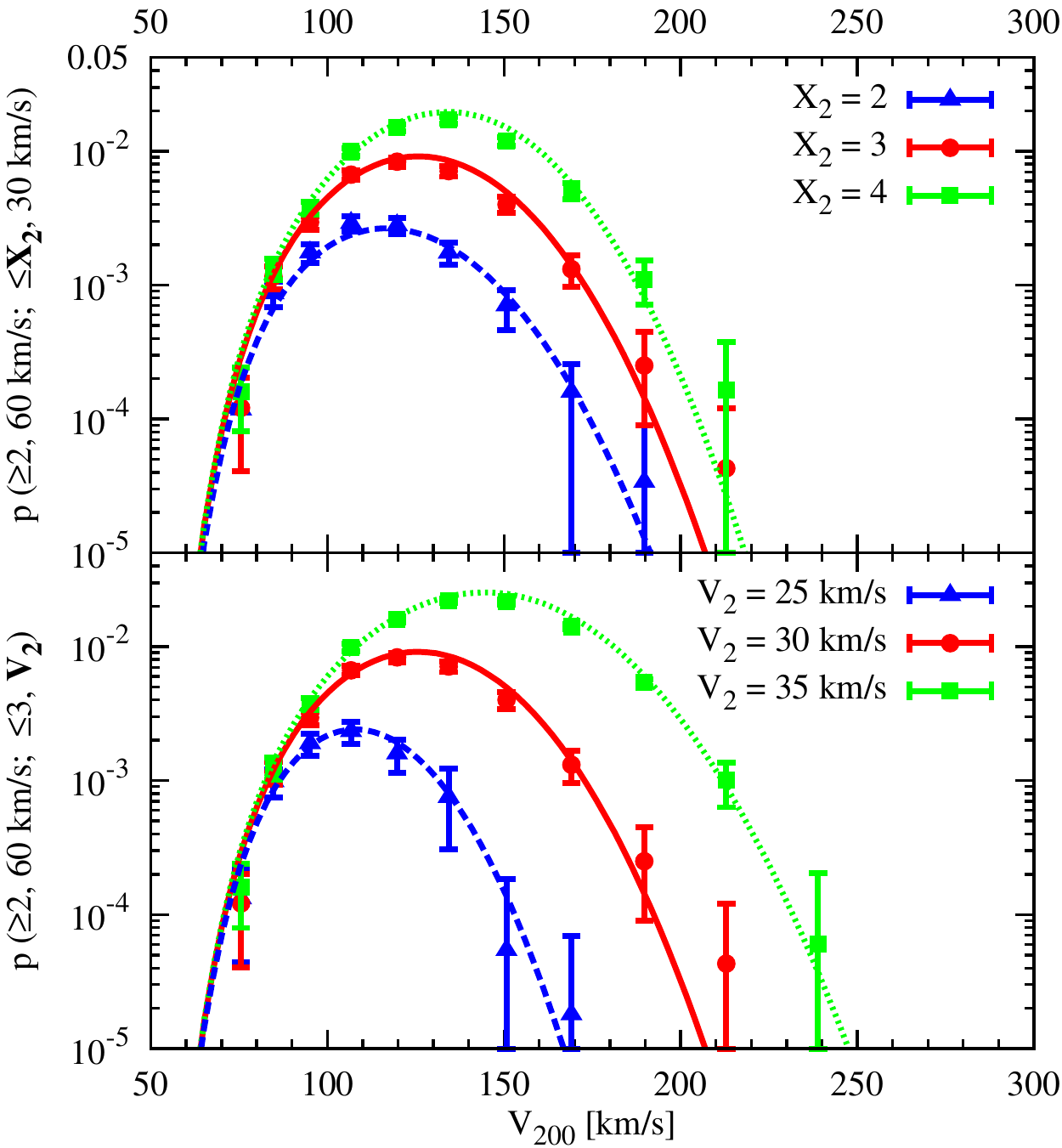}
     \caption{ Comparison of our theoretical model with results from
       the \MII{} simulations for the probability,
       $p({\ge}X_1,V_1;\;{\le}X_2,V_2)$, that a halo contains at least
       $X_1$ subhalos with $V_\rmn{max}\ge V_1$ and at most $X_2$
       substructures with $V_\rmn{max}\ge V_2$. We investigate
       departures from the default case,
       $p({\ge}2,60\kms;\;{\le}3,30\kms)$. In the top panel $X_{2}$ is
       varied while in the bottom panel $V_2$ is varied. The data
       points with bootstrap errors show the simulation results while the curves
       show the model predictions. }
     \label{fig:model_MW_like_subhalos}
\end{figure}

The subhalo number PDF diverges from a Poisson distribution for large
values of $\Ncum{}$ \citep[][\firstPaper{}]{Boylan-Kolchin2010} and
therefore our model gives only an approximate estimate of the true
probability. A more realistic description would involve the use of a
negative binomial distribution to characterise the probability for a
halo to have $l$ subhalos in the range $[\nu_2,\nu_1]$, but at the
expense of introducing an additional parameter. Since the deviation
from a Poisson distribution is small for $\nu\gsim0.15$ (\firstPaper{}), which defines
the region of interest here, we expect that our model gives a good
approximation to the probability of finding MW-like subhalo
populations.

In \reffig{fig:model_MW_like_subhalos} we compare the predictions of
our model to the results obtained from the \MII{} simulation. Since we
are interested in the probability of MW-like subhalo populations, we
explore a few representative examples close to this default case. In
the top panel we vary the number of subhalos, $X_2$, and in the right
panel the velocity threshold, $V_2$. For all cases we find that the
model predictions and the simulation data agree very well, showing
that our model gives a good approximation to the probability of
finding MW-like subhalo systems.


\section{Discussion}
\label{sec:MW_satellites:discussion}

The $V_\rmn{max}$ distribution of the Milky Way's most massive
satellites places strong constraints on the mass of the MW halo given
the prior hypothesis that \lcdm{} is the correct model. In this case,
the fact that the MW has only three satellites with
$V_\rmn{max}\ge30\kms$ (the two Magellanic Clouds and Sagittarius)
requires the virial mass of the MW halo to be $M_{200}
<1.4\times10^{12}\Msun$ at 90\% confidence; on the other hand, the
existence of the two Magellanic Clouds, which have
$V_\rmn{max}\ge60\kms$, requires $M_{200}> 1.0\times 10^{12}\Msun$, also
at 90\% confidence. This conclusion is consistent with some, but not
all, recent measurements of the MW mass 
\citep[][]{Battaglia2005,Smith2007,Xue2008,Guo2010,Watkins2010,Busha2011b,Gonzalez2014,Piffl2014,Diaz2014}.

These mass constraints were derived by treating the number of Galactic satellites with $V_\rmn{max}\ge60\kms$ and those with $V_\rmn{max}\ge30\kms$ as independent, which is clearly not the case. To overcome this, we defined halos with MW-like subhalo systems as those that have at most three satellites with $V_\rmn{max}\ge30\kms$, of which at least two have $V_\rmn{max}\ge60\kms$. In the simulation, the mass distribution of such halos is wider and shifted towards lower masses, suggesting a MW mass range of $0.25 \le M_{200}/(10^{12}\Msun) \le 1.4$ at $90\%$ confidence. \MCn{It is important to note that the low end of the $90\%$ confidence interval, $2.5\times10^{11}\Msun$, is likely ruled out by observations of the inner part of the Galactic halo. Using the fourth data release of the Radial Velocity Experiment \citep{Kordopatis2013}, \citet{Piffl2014} found that the MW halo mass within $180\kpc$ is $\ge{9\times10^{11}\Msun}$ at $90\%$ confidence \citep[][found similar lower bounds, albeit with larger uncertainties]{Smith2007,Xue2008,Gnedin2010,Deason2012}. This result could, in principle, be used as a prior for the kind of analysis we have carried out in this paper, along with other constraints coming from the orbital properties of the massive satellites \citep[e.g.][]{Busha2011b, Gonzalez2013} or the luminosity function of the nine bright ``classical'' dwarf spheroidal satellites \citep[][see also \citealt{Vera-Ciro2013}]{Kennedy2013}. }

Our results also confirm and extend the conclusion of \wang{} that the
``too-big-to-fail'' problem highlighted by
\cite{Boylan-Kolchin2011a,Boylan-Kolchin2012a} is not a problem for
the \lcdm{} model provided the MW halo mass is close to
$1\times 10^{12}\Msun$ rather than to the $\sim 2\times 10^{12}\Msun$
of the Aquarius halos used in the studies by Boylan-Kolchin et al.
Alternative solutions to the problem such as warm dark
matter \citep{Lovell2012}, self-interacting dark matter \citep{Vogelsberger2012}
or baryonic effects \citep{Brooks2013} are therefore
not required unless the mass of the MW halo can be shown to be
larger than $\sim 2\times 10^{12}\Msun$.

In our \lcdm{} simulations, halos with a $V_\rmn{max}$ distribution
similar to that of the MW, that is with at most three
satellites with $V_\rmn{max}\ge30\kms$, of which at least two have
$V_\rmn{max}\ge60\kms$, are rather rare as we have seen in 
\refsec{sec:MW_satellites:MW_typical}: at most $1\%$ of halos of 
any mass have satellite systems with this property.
This shows that the MW lies in the tail of the satellite distribution 
when analysing the cumulative satellite population at
 $V_\rmn{max,1}=30\kms$ and $V_\rmn{max,2}=60\kms$, which we call 
 ``the Galactic satellite gap". However, it is important to note that 
 this result does not necessarily imply a problem for the \lcdm{} paradigm.
To asses if the Galactic satellite gap represents a source of tension, 
we need to calculate what is the probability of finding such a gap in 
\lcdm{} haloes. For this, one needs to search for the presence of 
satellite gaps not only for $V_\rmn{max,1}=30\kms$ and 
$V_\rmn{max,2}=60\kms$, as we did here, but for all possible 
$V_\rmn{max,1}$ and $V_\rmn{max,2}$ combinations. It may be that 
satellite gaps are quite common, which would suggest that the 
Galactic satellite gap is a \lcdm{} prediction and not a cause of tension.


To assess the robustness of our conclusions we now explore their
sensitivity to various parameters required for this study.

\bigskip
\noindent{\em 1. Cosmological parameters }

\begin{figure}
     \centering
     \includegraphics[width=\linewidth,angle=0]{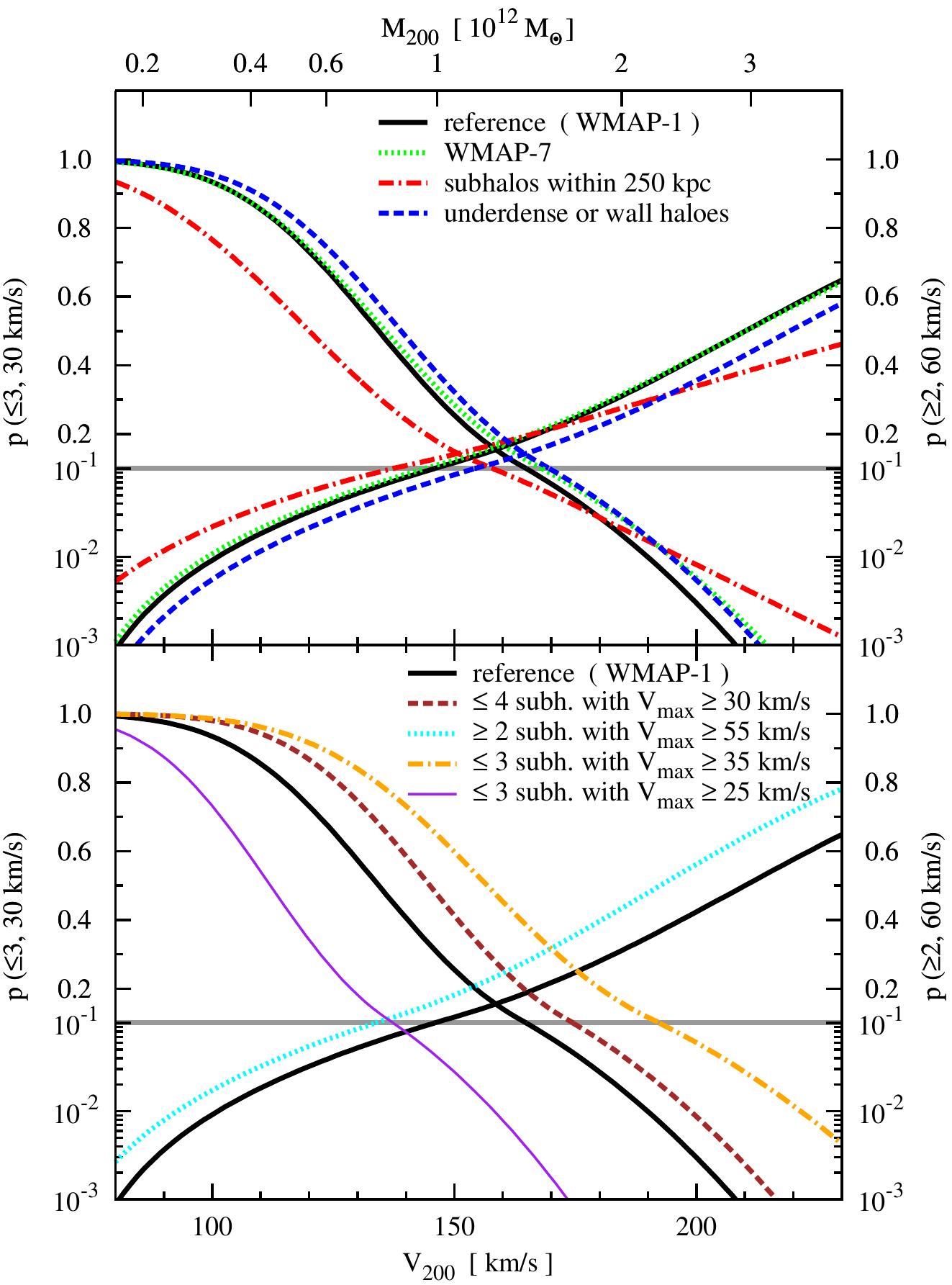}
     \caption{ The probability, $p({\le}3,30\kms)$, that
      a halo contains at most three subhalos with $V_{\rm max} \ge 30\kms$ (left y-axis) 
      and the probability, $p({\ge}2,60\kms)$ that a halo contains at least two subhalos 
      with $V_{\rm max}{\ge} 60\kms$ (right y-axis) as a function of halo
      virial velocity (lower x-axis) and virial mass (upper x-axis).
      The lines show the effect of changing some of the assumptions of the
      reference model studied until now. The solid curves show the reference case of WMAP-1
      cosmological parameters and substructures found with $R_{200}$ from the host halo
      centre (as in \reffigS{fig:MW_constraints_1}{fig:MW_constraints_2}).
      \textit{Top}: results for WMAP-7 cosmological parameters (dotted green), predictions
      when subhalos within \MCn{a distance of $250\kpc$ from} the host centre are considered (dashed-dotted red) 
      and the effect of large scale environment by considering only host halos found in 
      underdense or wall regions (dashed blue).
      \textit{Bottom}: outcome of assuming that \MCn{the MW has $4$ (instead of $3$; dashed brown) satellites with $V_\rmn{max}\ge30\kms$},
      effect of assuming the MCs have $V_\rmn{max}\ge55\kms$ (instead of $60 \kms$; dotted cyan)
      and the results of assuming that the MW has at most three satellites with velocity 
      threshold $V_\rmn{max}\ge35\kms$ (dashed-dotted golden) and $V_\rmn{max}\ge25\kms$ (thin solid purple)
      respectively (instead of $30 \kms$). The horizontal gray line shows the 10\% level.
     }
     \label{fig:MW_mass_constraints_1}
\end{figure}

\begin{figure}
     \centering
     \includegraphics[width=\linewidth,angle=0]{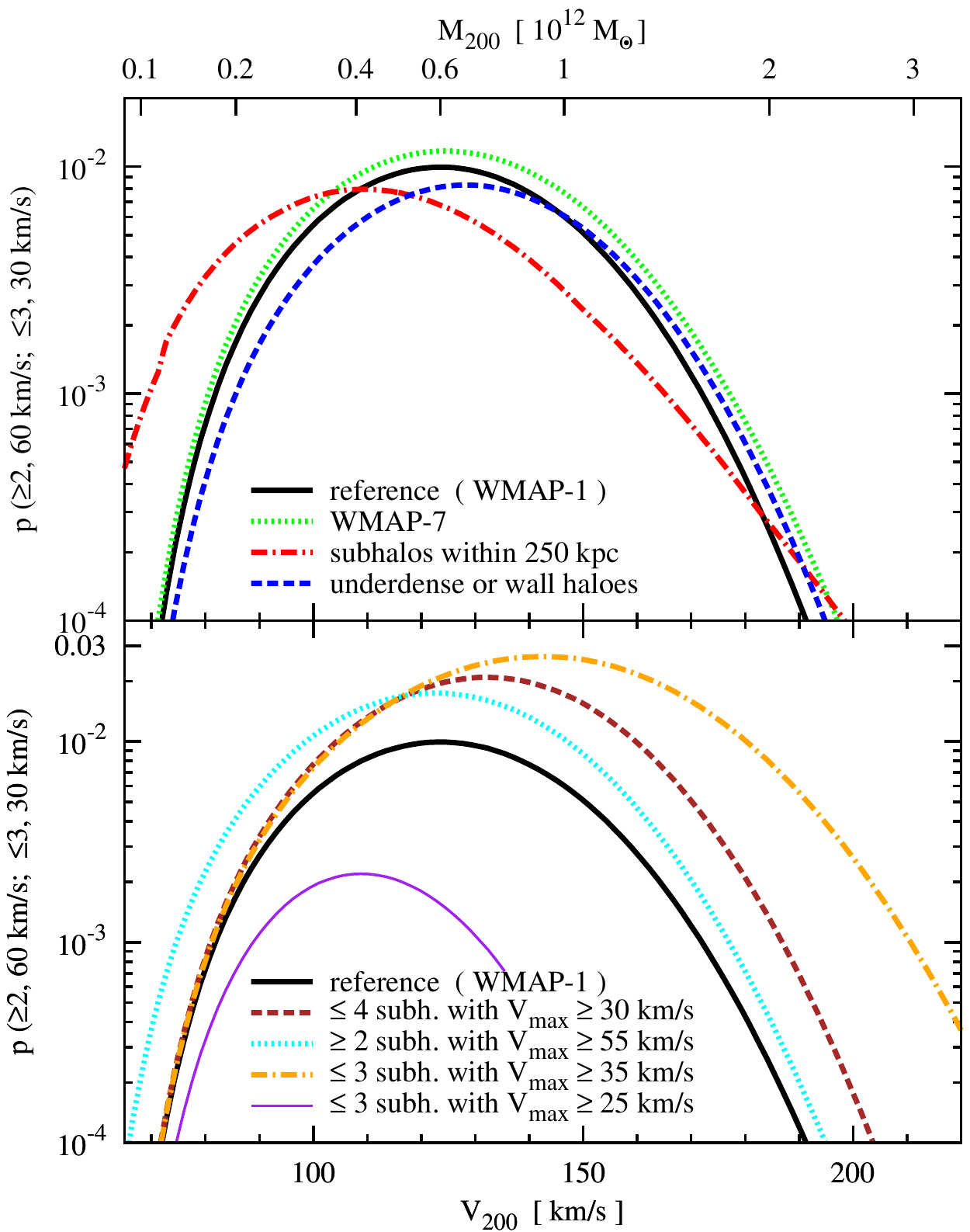}
     \caption{ The probability that a halo contains a
       MW-like subhalo system as a function of halo virial velocity
      (lower x-axis) or virial mass (upper x-axis). 
      The different lines show the effect of changing some of the assumptions of the
      reference model studied until now.  
      We explore the same variations from the reference model as in \reffig{fig:MW_mass_constraints_1}.
       }
     \label{fig:MW_mass_constraints_2}
\end{figure}

The results presented here are based on the \MII{} that
assumed WMAP-1 values for the cosmological parameters. The main
difference between these and more recent measurements from WMAP-7
\citep{Komatsu2011} or the Planck satellite \citep{Planck2013_XVI} is
a lower value of $\sigma_8$. \firstPaper{} found that lowering the
value of $\sigma_8$ from the WMAP-1 value of 0.9 to the WMAP-7 value
of 0.8 results in a slightly lower number of substructures.  This
translates into a slightly different allowed range for the Milky Way
halo mass, as seen from \reffigS{fig:MW_mass_constraints_1}{fig:MW_mass_constraints_2}.
The probability of finding a MW-like subhalo
population assuming WMAP-7 parameters (dotted green line in
\reffig{fig:MW_mass_constraints_2}) increases slightly and the peak shifts
towards higher masses, but the overall difference is very small. 
For convenience, we summarized in \reftab{tab:probability_peak_values} the variations in both the mass estimate and peak height.

\medskip
\noindent{\em 2. Maximum distance used to identify satellites}

Our analysis so far has been based on substructures found within the
virial radius, $R_{200}$, of the host halo centre. For halo masses of
$10^{12}\Msun$ \MCn{and lower} this distance corresponds to ${\lsim}200\kpc$ and it is \MCn{significantly}
smaller than the distances of the outermost known satellites of the
MW, such as Leo I, which lies at ${\sim}250\kpc$ from the halo centre 
\citep{Karachentsev2004}. To assess the impact of our choice of
radius, we repeated the analysis including subhalos located within a \MCn{fixed distance of $250\kpc$ from the host center, independently on the host mass (see \refappendix{appendix_subhalos_250kpc} for details).} 
The results are shown in \reffigS{fig:MW_mass_constraints_1}{fig:MW_mass_constraints_2} as the
dotted-dashed red curve that can be compared with the solid curve for our
default case. \MCn{The difference arises because $R_{200}<250\kpc$ for halo masses below $1.5\times10^{12}\Msun$, which are of interest for our comparison.}
Since the number of massive substructures increases
rapidly with the value of the limiting radius, it becomes more
difficult to find halos with at most three $V_{\rm max}{\ge} 30\kms$
subhalos and this has the effect of lowering the upper limit on the MW
halo mass. On the other hand, it becomes easier to find at least
two substructures with $V_{\rm max}\ge 60\kms$ and this has the effect
of also lowering the lower limit on the MW halos mass. 
The net effect is to shift the allowed mass range to lower values, $0.15 \le M_{200}/(10^{12}\Msun) \le 1.2$ at 90\% confidence, \MCn{and to reduce the peak probability of finding a MW-like subhalo system}.

\bigskip 
\noindent{\em 3. Velocity thresholds}

A key ingredient of our analysis are the two velocity thresholds
that we use to characterise the MW satellites: $30 \kms$ for the
threshold above which there should be no more that three subhalos and
$60 \kms$ for the threshold above which there should be at least two
subhalos. Increasing the first of these thresholds to $35 \kms$ has the effect of
weakening the upper limit on the MW halo mass to $M_{200} \lsim 2.1\times10^{12}\Msun$
(90\% confidence; see dashed-dotted golden line in 
\reffigS{fig:MW_mass_constraints_1}{fig:MW_mass_constraints_2}). 
However, decreasing this threshold to $25 \kms$ \citep[as 
suggested by][]{Boylan-Kolchin2012a} has a more dramatic effect 
(thin-solid purple curve in \reffigS{fig:MW_mass_constraints_1}{fig:MW_mass_constraints_2}), 
giving a mass range of 
$0.19 \le M_{200}/(10^{12}\Msun) \le 0.82$ at $90\%$ confidence.
The likelihood of finding  MW-like subhalo systems for these values of
the thresholds varies by factors of a few from the reference case: for
the $25\kms$ threshold only ${\sim}0.3\%$ of \lcdm{} halos have such
subhalo systems {while for a $35\kms$ threshold the probability increases to ${\sim}3\%$.

Regarding the second velocity threshold, the uncertainties of the best available
measurements of the Small Magellanic Cloud's rotation velocity are consistent with a value
of $V_\rmn{max}=55\kms$ \citep{Kallivayalil2013}. This
change has the effect of slightly weakening the lower limit on the halo mass 
(dotted cyan curve in \reffigS{fig:MW_mass_constraints_1}{fig:MW_mass_constraints_2}).
The probability of finding a MW-like subhalo population 
increases to $1.7\%$, but the peak position remains unchanged.

In conclusion, our results are most sensitive to the first velocity
threshold of $30\kms$, which is also the one most prone to measurement 
and modelling uncertainties since it is derived by studying the 
kinematics of the nine bright ``classical'' dwarf spheroidal satellites.

\begin{table*}
    \small
    \centering
    \caption{ The sensitivity of the MW mass estimation on the various parameters used in our study.
       It shows the MW mass range, at 90\% confidence, as inferred for the various 
       cases explored in \reffig{fig:MW_mass_constraints_1} (third column) 
       and \reffig{fig:MW_mass_constraints_2} (fourth column). We also give the 
       peak value (sixth column) and the halo mass at the peak position (fifth column) for each of the datasets 
       shown in \reffig{fig:MW_mass_constraints_2}. }
    \label{tab:probability_peak_values}
    \begin{tabular}{llcccc}
        \hline \hline
        Dataset & \parbox[][][c]{0.13\textwidth}{Representation in \reffigS{fig:MW_mass_constraints_1}{fig:MW_mass_constraints_2}} 
            &  \multicolumn{2}{c}{ \parbox[][][c]{0.21\textwidth}{\centering MW mass limits [$\times 10^{12}\Msun$] (90\% confidence)} } 
            & \parbox[][][c]{0.16\textwidth}{\centering mass at peak position [$\times 10^{12}\Msun$]} &  \parbox[][][c]{0.08\textwidth}{\centering peak value [$\%$]}  \\
        \hline
        WMAP-1 reference result  &   solid black      
            &   $1.0 - 1.4$ &   $0.25 - 1.4$ &   $0.61$ &   $1.0$   \\
        WMAP-7 cosmology         &  dotted green                                                    
            &   $1.0 - 1.6$ &   $0.26 - 1.5$ &   $0.64$ &   $1.2$   \\
        subhalos within $250\kpc$  & dashed-dotted red                             
            &  $0.83 - 1.2$ &   $0.15 - 1.2$ &   $0.42$ &   $0.80$   \\
        underdense or wall halos & dashed blue                                           
            &   $1.2 - 1.6$ &   $0.28 - 1.5$ &   $0.68$ &   $0.83$   \\
        ${\le}4$ subhalos with $V_\rmn{max}\ge30\kms$    & dashed brown                                                         
            &   $1.0 - 1.7$ &   $0.29 - 1.5$ &   $0.74$ &   $2.1$   \\
        ${\ge}2$ subhalos with $V_\rmn{max}\ge55\kms$    & dotted cyan                              
            &  $0.77 - 1.4$ &   $0.23 - 1.3$ &   $0.60$ &   $1.7$   \\
        ${\le}3$ subhalos with $V_\rmn{max}\ge35\kms$    & dashed-dotted golden                
            &   $1.0 - 2.3$ &   $0.30 - 2.1$ &   $0.93$ &   $2.7$   \\
        ${\le}3$ subhalos with $V_\rmn{max}\ge25\kms$    & thin-solid purple                
            &   $-$         &  $0.19 - 0.82$ &   $0.38$ &  $0.28$   \\
        \hline
    \end{tabular}
\end{table*}

\bigskip 
\noindent{\em 4. \MCn{Incompleteness of MW satellites}}

\MCn{The sample of MW satellites is possibly incomplete, with the recent study of \citet{Yniguez2014} suggesting that around $10$ dwarf spherodial satellites await discovery in the area left unexplored by the Sloan Digital Sky Survey. It is possible, though unlikely, that one or more of these undiscovered satellites could have $V_{\rmn{max}}\ge30\kms$. In addition, recent dynamical modelling of the Sculptor dwarf spheroidal galaxy performed by \citet{Strigari2014} has found that the observational data allow for a maximum circular velocity up to ${\sim}35\kms$. The presence of an additional massive satellite would have the effect of weakening the upper limit on the MW halo mass to $M_{200} \lsim 1.5\times10^{12}\Msun$ ($90\%$ confidence) and increasing the probability of finding a MW-like subhalo system (dashed brown curve in \reffigS{fig:MW_mass_constraints_1}{fig:MW_mass_constraints_2}). }

\bigskip 
\noindent{\em 5. Environmental effects}

Recent studies have shown that the number of substructures depends on
the large scale environment, with halos in lower density regions
having fewer subhalos \citep{Ishiyama2008,Busha2011,Croft2012}.  This
trend has been further quantified by Cautun et al. (in prep.) who find
that this effect is significant only for halos in the most underdense
regions and for those residing in the sheets of the cosmic web. These
halos have, on average, $10-20\%$ fewer substructures than the
population as a whole, and the deficiency is larger for more massive
subhalos. Environmental effects of this kind may play a role in our
galaxy since both observational and theoretical considerations suggest
that the Local Group lies within a large-scale sheet
\citep{Tully1988,Pasetto2009,Aragon-Calvo2011}.

To assess the importance of this kind of environmental effect, we
have applied NEXUS \citep[][]{Cautun2013}, a morphological environment identification method,
to count the substructures of halos that reside in
different environments. The paucity of the most massive subhalos
within wall halos has the effect of increasing both the lower and
upper limits on the allowed MW halo mass (dashed blue curve in 
\reffigS{fig:MW_mass_constraints_1}{fig:MW_mass_constraints_2}) so that 
the allowed interval shifts to ${\sim}10\%$ higher halo masses 
(see \reftab{tab:probability_peak_values} for details).
The probability of finding a MW-like subhalo system
is only slightly lowered.

\bigskip 
\noindent{\em \MCn{6. Baryonic effects}}

\MCn{Baryonic processes are known to affect the mass function and inner structure of halos, especially at the low mass end. For example,
\cite{Sawala2013,Sawala2014} have shown that baryonic
effects in simulations of galaxy formation cause halos with mass
$\lsim 10^{11}\Msun$ to grow at a reduced rate compared to their
counterparts in a dark matter only simulation. 
Baryonic processes also affect the maximum circular velocity of galactic satellites, especially dwarf spheroidal galaxies \citep[e.g.][]{Zolotov2012,Brooks2014}, which can have important implications for our study. The inclusion of baryons does not affect the maximum circular velocity of massive satellites with $V_{\rmn{max}}\sim60\kms$, but it does lead to an average ${\sim}10\%$ reduction in the maximum circular velocity of satellites with $V_{\rmn{max}}\lsim30\kms$ (Sawala et al. in prep., private communication). These results are based on a comparison of matched satellites between dark matter only and hydrodynamic simulations, in a set of $24$ distinct MW mass halos \citep[The suite of simulations is described in][]{Sawala2014b}. Thus, dwarf spheroidals that have $V_{\rmn{max}}\lsim30\kms$ correspond to subhalos that, in the dark matter only simulations, have a factor of ${\sim}1.1$ higher maximum circular velocity. This can be easily incorporated into our analysis by changing the condition of finding at most three subhalos with $V_{\rmn{max}}\ge30\kms$ to the conditions of finding at most three subhalos with $V_{\rmn{max}}\ge34\kms$. This weakens the upper limit to the MW halo mass to $M_{200} \lsim 1.9\times10^{12}\Msun$ ($90\%$ confidence; for clarity we do not show this curve in \reffigS{fig:MW_mass_constraints_1}{fig:MW_mass_constraints_2} but its position can be easily estimated by comparing to the dashed-dotted golden line corresponding to $V_{\rmn{max}}\ge35\kms$).  }


\section{Summary}
\label{sec:MW_satellites:conclusion}

We have employed the $V_\rmn{max}$ distribution of satellites in the
MW to set lower and upper limits to the virial mass of the Galactic
halo and to find how likely the MW satellite system is under the
assumption that \lcdm{} is the correct model for 
cosmic structure formation. The upper limit comes from requiring that the
MW should have at most three subhalos with $V_\rmn{max}\ge30\kms$; the
lower limit comes from requiring that the MW should have at least two
subhalos with $V_\rmn{max}\ge60\kms$. The first of these requirements
is necessary to avoid the TBTF problem highlighted by
\cite{Boylan-Kolchin2011a,Boylan-Kolchin2012a}, while the second stems
from the observation that massive satellites like the MCs are rare \citep{Liu2011,Guo2011,Lares2011}.

Our analysis is based on over $10^4$ halos from the Millennium-II
simulation. To achieve the required dynamic range, we use an 
extrapolation method devised by \firstPaper{} that allows us to count
subhalos down to $V_\rmn{max}\sim15\kms$. 
In a first step we estimate lower and upper bounds to the MW halo mass by treating the number of satellites with $V_\rmn{max}\ge60\kms$ and those with $V_\rmn{max}\ge30\kms$ as independent. The former requirement implies a MW mass of $M_{200}\ge1.0\times10^{12}\Msun$ while the latter condition indicates that $M_{200}\le1.4\times10^{12}\Msun$, with both limits given at $90\%$ confidence. When requiring that host haloes have a $V_\rmn{max}$ distribution similar to that of the MW, that is with at most three satellites with $V_\rmn{max}\ge30\kms$, of which at least two have $V_\rmn{max}\ge60\kms$, the allowed mass range becomes $0.25 \le M_{200}/(10^{12}\Msun) \le 1.4$ ($90\%$ confidence). 

We also find that the $V_\rmn{max}$ distribution of the massive subhalos of the MW, as defined by the number of satellites with $V_\rmn{max}\ge30\kms$ and those with $V_\rmn{max}\ge60\kms$,
is quite rare in \lcdm{} simulations, with at most ${\sim}1\%$ of
halos of any mass having a similar distribution. This might be signalling a
tension between the \lcdm{} model and observations of the MW satellites, 
but it is not clear that constructing a solid statistical analysis 
on such an {\em a posteriori} argument is possible without
a detailed analysis of the frequency of gaps as a function of the
threshold values of $V_\rmn{max}$.

Our conclusion regarding the rarity of the MW subhalo system does not
vary significantly when we vary the parameters of our model. However,
the allowed mass for the MW halo is sensitive to uncertainties in the 
parameters we use, especially in the $V_\rmn{max}=30\kms$ threshold 
that is derived from the kinematics of the nine bright ``classical'' 
dwarf spheroidal satellites. Thus, as pointed out by \wang{} and \firstPaper{},
the TBTF problem is easily avoided if the MW halo has a
relatively low mass, certainly within the range of current
measurements. However, our study highlights the importance for
cosmology of obtaining robust and reliable measurements of the mass of
the MW's halo.

\section*{Acknowledgements}

We thank the referee for their useful comments that have improved this paper. We are also grateful to Shaun Cole, Vincent Eke, Douglas Finkbeiner, Julio Navarro, Till Sawala and Andrew Pontzen for helpful discussions and suggestions.
This work was supported in part by ERC Advanced Investigator grant COSMIWAY 
[grant number GA 267291] and the Science and Technology Facilities Council 
[grant number ST/F001166/1, ST/I00162X/1]. RvdW acknowledges support by the John 
Templeton Foundation, grant number FP05136-O. WAH is also supported by the Polish
National Science Center [grant number DEC-2011/01/D/ST9/01960]. The simulations used 
in this study were carried out by the Virgo consortium for cosmological simulations. 
Additional data analysis was performed on the Cosma cluster at ICC in Durham and 
on the Gemini machines at the Kapteyn Astronomical Institute in Groningen. 

This work used the DiRAC Data Centric system at Durham University, 
operated by ICC on behalf of the STFC DiRAC HPC Facility (www.dirac.ac.uk). 
This equipment was funded by BIS National E-infrastructure capital 
grant ST/K00042X/1, STFC capital grant ST/H008519/1, and STFC DiRAC 
Operations grant ST/K003267/1 and Durham University. DiRAC is part 
of the National E-Infrastructure. Data from the Millennium/Millennium-II 
simulation is available on a relational database accessible from 
http://galaxy-catalogue.dur.ac.uk:8080/Millennium . This research was carried out with the support of the ``HPC Infrastructure for Grand Challenges of Science and Engineering'' Project, co-financed by the European Regional Development Fund under the Innovative Economy Operational Programme.


\newcommand{\jcap}{Journal of Cosmology and Astroparticle Physics}
\bibliographystyle{mn2e}
\bibliography{references_bib}

\appendix


\section{The probability of finding MW-like satellites}
\label{appendix_model_MW_like_subhalos}
Here we give a detailed description of the model that we use to
predict the probability, $p({\ge}X_1,V_1;\; {\le}X_2,V_2)$, that
a halo contains at least $X_1$ subhalos with $V_\rmn{max}\ge V_1$ and
at most $X_{2}$ substructures with $V_\rmn{max}\ge V_2$, where $V_1\ge
V_2$. For simplicity, we use the notation
\begin{equation}
    \mathcal{P} = p({\ge}X_1,V_1;\; {\le}X_2,V_2)
\end{equation}
and we take $X_2\ge X_1$. The case $X_2< X_1$ is trivial since the probability is zero.

In the  first instance we restrict attention to host halos with
virial velocity, $V_{200}$. Using the notation, 
\begin{equation}
    \nu_1 = \frac{V_1}{V_{200}} \mbox{\hspace{0.5cm} and \hspace{0.5cm}} 
    \nu_2 = \frac{V_2}{V_{200}} \; ,
    \label{eq:appendix:nu_1_nu_2}
\end{equation}
the probability $\mathcal P$ reduces to finding all the halos with
$V_{200}$ that contain at least $X_1$ subhalos with $\nu\ge \nu_1$
and at most $X_2$ subhalos with $\nu\ge \nu_2$. At $\nu_2$ there are, 
on average,
\begin{equation}
    \Delta N = \overline{N}({>}\nu_2) - \overline{N}({>}\nu_1)
\end{equation}
more substructures per halo than at $\nu_1$, where
$\overline{N}({>}\nu_1)$ and $\overline{N}({>}\nu_2)$ are the mean
subhalo counts at those two velocity ratios. We make the assumption
that these subhalos with $\nu\in[\nu_2,\nu_1]$ are distributed among
the host population according to a Poisson distribution with mean
$\Delta N$ that is independent on the number of substructures at
$\nu_1$. Therefore, a halo has a probability, 
\begin{equation}
    P_\rmn{Poisson}(l,\Delta N) = \frac{\Delta N^l}{l!} e^{-\Delta N}, 
    \label{eq:appendix:delta_N}
\end{equation}
of having $l$ subhalos with $\nu\in[\nu_2,\nu_1]$. The same halo has probability
\begin{equation}
    P_\rmn{Poisson}({\le}i,\Delta N) = \sum_{l=0}^i \frac{\Delta N^l}{l!} e^{-\Delta N}
    \label{eq:appendix:poisson_cumulative}
\end{equation}
of having  at most $i$ substructures in the range $[\nu_2,\nu_1]$.

The only halos that contribute to $\mathcal P$ are those that have
between $X_1$ and $X_2$ substructures with $\nu\ge\nu_1$. Let us
select such a halo containing $k\in[X_1,X_2]$ subhalos with
$\nu\ge\nu_1$. This halo can contribute to $\mathcal P$ only if it has
at most $X_2$ substructures with $\nu\ge\nu_2$ and therefore it 
can have at most $X_2-k$ subhalos in the range $[\nu_2,\nu_1]$. The
probability that it satisfies this condition is given by
\eq{eq:appendix:poisson_cumulative} with $i=X_2-k$.

The quantity, $\mathcal P$, is given by the fraction of halos with $k$
substructures at $\nu\ge\nu_1$ times the probability that they contain
less than $X_2-k$ subhalos in the range $[\nu_2,\nu_1]$, summed over
$k$. Therefore, we have, 
\begin{equation}
    \mathcal{P} = \sum_{k=X_1}^{X_2} P(k|r({>}\nu_1),s({>}\nu_1)) \;  P_\rmn{Poisson}({\le}X_2{-}k,\Delta N) \,,
    \label{eq:appendix:MW_like_subhalos}
\end{equation}
where $P(k|r({>}\nu_1),s({>}\nu_1))$ is the negative binomial
distribution that gives the probability that a halo has $k$
substructures with $\nu>\nu_1$ (see Eqs. \ref{eq:MW_satellites:NBD}
and \ref{eq:MW_satellites:NBD_parameters}). The probability,
$\mathcal{P}$, is a function of halo virial velocity, or equivalently,
halo mass, through the dependence of $r$ and $s$ on $\nu_1$ as well as
the variation of $\Delta N$ with $\nu_1$ and $\nu_2$.


\section{\MCn{The subhalo abundance within a fixed physical radius}}
\label{appendix_subhalos_250kpc}
\MCn{To compute the subhalo abundance within a fixed physical radius we make use of the universality of $\Ncum{}$ with host halo mass. This approximation is valid when $\Ncum{}$ is measured within a distance $fR_{200}$, with $f$ a multiplication factor. This is illustrated in \reffig{fig:statistics_mass_dependence} for a value of $f=1$. We have checked that the universality still applies, to within ${\sim}20\%$, for the mass range $1\times10^{11}\Msun\le M_{200} \le 1\times10^{13}\Msun$, for values of $f$ in the range $0.5\le f \le 3.0$. }

\MCn{Computing the subhalo abundance within a fixed physical radius, $R$, is equivalent to a distance, $fR_{200}$, with
\begin{equation}
    f \equiv \frac{R}{R_{200}} \;.
\end{equation}
Since $R_{200}$ is a function of mass, the multiplication factor, $f$, is itself a function of halo mass, with $f$ decreasing with increasing halo mass. We computed the subhalo abundance within a distance of $fR_{200}$ for a set of $f$ values in the range $0.57$ to $2.7$, which corresponds to a fixed distance of $R=250\kpc$ spanning the mass range $1\times10^{11}\Msun \le M_{200} \le 1\times10^{13}\Msun$. The $f$ values were selected to give nine equally spaced bins in $M_{200}$. Following this, the abundance of subhalos at a given halo mass was found using a linear interpolation between the results for the two closest values of $f$ corresponding to that mass value. }

\end{document}